\documentclass[10pt, aps, prl, twocolumn, superscriptaddress, floatfix, showpacs, longbibliography]{revtex4-1}

\usepackage{graphicx}
\usepackage{amsfonts}
\usepackage{amssymb}
\usepackage{amsmath}
\usepackage{txfonts}
\usepackage{lipsum}
\usepackage{color}
\usepackage{wasysym}
\usepackage[colorlinks=true, allcolors=blue]{hyperref}
\usepackage{bbold}
\usepackage{etoolbox} 
\usepackage[normalem]{ulem}
\usepackage{mathtools} 
\usepackage{multirow} 
\usepackage{hhline}
\usepackage{mathrsfs} 
\usepackage{soul}

\usepackage{letltxmacro}
\LetLtxMacro{\oldsqrt}{\sqrt}
\renewcommand{\sqrt}[2][\mkern8mu]{\mkern-6mu\mathop{}\oldsqrt[#1]{#2}}

\usepackage{url}
\definecolor{indigo(dye)}{rgb}{0.0, 0.25, 0.42}
\usepackage{placeins}

\begin{document}

\title{Charge Density Wave Ordering in NdNiO$_2$: Effects of Multiorbital Nonlocal Correlations}

\author{Evgeny A. Stepanov}
\affiliation{CPHT, CNRS, {\'E}cole polytechnique, Institut Polytechnique de Paris, 91120 Palaiseau, France}
\affiliation{Coll\`ege de France, Universit\'e PSL, 11 place Marcelin Berthelot, 75005 Paris, France}

\author{Matteo Vandelli}
\affiliation{The Hamburg Centre for Ultrafast Imaging, Luruper Chaussee 149, 22761 Hamburg, Germany}

\author{Alexander I. Lichtenstein}
\affiliation{Institut f{\"u}r Theoretische Physik, Universit{\"a}t Hamburg, Notkestra{\ss}e 9 , 22607 Hamburg, Germany}
\affiliation{The Hamburg Centre for Ultrafast Imaging, Luruper Chaussee 149, 22761 Hamburg, Germany}

\author{Frank Lechermann}
\affiliation{Institut f{\"u}r Theoretische Physik III, Ruhr-Universit{\"a}t Bochum, D-44780 Bochum, Germany}

\begin{abstract}
In this work, we investigate collective electronic fluctuations and, in particular, the possibility of the charge density wave ordering in an infinite-layer NdNiO$_2$.
We perform advanced many-body calculations for the \emph{ab-initio} three-orbital model by taking into account local correlation effects, nonlocal charge and magnetic fluctuations, and the electron-phonon coupling.
We find that in the considered material, electronic correlations are strongly orbital- and momentum-dependent. Notably, the charge density wave and magnetic instabilities originate from distinct orbitals.
In particular, we show that the correlation effects lead to the momentum-dependent hybridization between different orbitals, resulting in the splitting and shifting of the flat part of the Ni-$d_{z^2}$ band.
This strong renormalization of the electronic spectral function drives the charge density wave instability that is related to the intraband Ni-$d_{z^2}$ correlations. 
Instead, the magnetic instability stems from the Ni-$d_{x^2-y^2}$ orbital, which remains half-filled through the redistribution of the electronic density between different bands even upon hole doping.
Consequently, the strength of the magnetic fluctuations remains nearly unchanged for the considered doping levels.
We argue that this renormalization is not inherent to the stoichiometric case but can be induced by hole doping.
\end{abstract}

\maketitle

\section{Introduction}

The discovery of superconductivity (SC) in layered nickel-oxide compounds, has been one of the most thrilling research findings in the recent years. It started off from revealing SC in thin-films of infinite-layer nickelates upon hole doping $\delta$ with ${T_{\rm c}\sim 15}$\,K~\cite{PhysRevLett.125.027001, PhysRevLett.125.147003, osada20, osada21, zeng22}, and afterwards also in low-valence multilayer nickelates~\cite{pan21}. As a last key development, high-$T_{\rm c}$ SC with ${T_{\rm c}\sim 80}$\,K has been uncovered in the pressurized bilayer compound La$_3$Ni$_2$O$_7$~\cite{sun23}. 
Unconventional SC is often escorted by other electronic orders in an intriguing phase diagram, but the usual suspect of magnetic-ordering kind remains elusive in these nickelates. However, there have been several experimental reports of charge-density wave (CDW) ordering in the thin-films of rare-earth infinite-layer nickelates, originating at stoichiometry and stable up to ambient temperatures. This detected CDW order is associated with an in-plane incommensurate wave vector ${{\bf q}_{\rm CDW}\sim (0.3,0)}$ and involves Ni$(3d)$ and rare-earth$(5d)$ degrees of freedom~\cite{rossi22, krieger22, tam22, ren2023}. 
But recently, a debate has started about the role of the substrate and of possible impurity phases in view of the CDW findings in these thin-film systems~\cite{https://doi.org/10.1002/smll.202304872, parzyck2023, pelliciari2023, tam2023reply, hayashida2024investigation}.

The theoretical modelling of the correlated electronic structure of superconducting nickelates is also not yet settled. 
The physics of these systems is governed by a complex interplay between strong local Coulomb correlations, spatial collective electronic fluctuations, orbital degrees of freedom, and a non-trivial band structure.
Density functional theory (DFT) calculations for infinite-layer nickelates~\cite{PhysRevB.100.205138, PhysRevB.101.081110} describe a nearly half-filled Ni-$d_{x^2-y^2}$-dominated band at the Fermi level, with additional electron pockets from a self-doping (SD) band (solid black lines in Fig.~\ref{fig:LDA_DMFT}). Around the $\Gamma$ point, the latter is based on significant contribution from Ni-$d_{z^2}$ and Nd-$d_{z^2}$.
In addition, at ${k_{z}=\pi}$ momentum the Ni-$d_{z^2}$ band displays a flat part, which may trigger various collective electronic instabilities if appearing at the Fermi energy~\cite{PhysRevB.64.165107, PhysRevB.67.125104, RevModPhys.84.299}.
Until now, even the most advanced numerical calculations could not account for all these important effects simultaneously.
Especially for these low-valence $3d^{9-\delta}$ systems of infinite-layer and multilayer kind, there is thus an ongoing debate concerning the picture most-effectively describing the low-energy physics: a single dominant Ni-$d_{x^2-y^2}$ orbital~\cite{PhysRevB.101.060504, PhysRevB.101.020501, PhysRevX.10.021061, PhysRevB.101.241108, PhysRevB.102.100501, kitatani2020, PhysRevX.11.011050, PhysRevResearch.3.013261, gu2020substantial, PhysRevB.106.134504, PhysRevLett.124.207004} versus a mainly Ni-$e_g$ multi-orbital character~\cite{PhysRevB.101.041104, PhysRevB.101.081110, PhysRevX.10.041002, PhysRevX.10.041047, PhysRevLett.126.127401, PhysRevLett.129.077002, abadi2024electronic}.
Neglecting spatial collective electronic fluctuations makes possible the {\it state-of-the-art} dynamical mean-field theory (DMFT)~\cite{RevModPhys.68.13} calculations for an {\it ab-initio} multi-orbital model. On the other hand, a single-orbital description allows for accounting for the spatial correlation effects that are missing in DMFT~\cite{kitatani2020, worm2023spin}.
The correlation strength in different model calculations varies from a weakly-to-moderately metallicity up to (orbital-selective) Mott-critical regimes. In view of a possible CDW ordering, calculations which in the end focus on a dominant Ni-$d_{x^2-y^2}$ low-energy degree of freedom may indeed give reason for the experimentally detected incommensurate order~\cite{PhysRevB.107.165103, chen23}.

In this work, the possibility of a CDW instability in the infinite-layer nickelate NdNiO$_2$, at stoichiometry as well as upon hole doping, is examined by a model Hamiltonian study based on its realistic low-energy electronic structure. 
This modeling goes beyond the sole Ni-$d_{x^2-y^2}$-based physics, allowing for multi-orbital Ni-$e_g$ processes interplaying with the SD band.
Furthermore, our numerical approach accounts for the combined effect of strong local electronic correlations, nonlocal collective electronic fluctuations, and phonon degrees of freedom. 
This surpasses any theoretical description ever performed for this class of materials.

Our results demonstrate that accounting for both spatial electronic correlations and orbital degrees of freedom is important. 
We find that the nonlocal electronic correlations in NdNiO$_2$ result in a substantial momentum- and orbital-dependent renormalization and hybridization of all bands considered in our {\it ab-initio} model, which cannot be captured neither on the basis of local theories nor within a single-band framework.
As a main result, we show that the CDW ordering in the infinite-layer NdNiO$_2$ originates from the intraband correlations within the Ni-$d_{z^2}$ orbital.
Upon the renormalization, the flat part of this band, which originally lies below the Fermi level, splits into two parts. 
One part moves toward the Fermi energy and can even appear above the Fermi level. 
The other part moves in the opposite direction and hybridizes with the lower Hubbard band of the Ni-$d_{x^2-y^2}$ orbital.
In addition, we find that electronic correlations enlarge the electron pocket around the $\Gamma$ point, which corresponds to the hybridized Ni-$d_{z^2}$ and SD bands. 
This enhancement eventually leads to the nesting of the Fermi surface and to the CDW instability.

We argue that such momentum-dependent renormalization of the electronic spectral function is unlikely to occur at stoichiometry, in agreement with recent experimental works that conclude on the absence of the CDW instability in the undoped case~\cite{https://doi.org/10.1002/smll.202304872, parzyck2023, hayashida2024investigation}.
However, we further demonstrate that the CDW instability can be induced upon hole doping the system.
Remarkably, the estimated critical value of the electronic density for the CDW phase transition agrees well with the position of the dip in the superconducting dome observed in the hole doped NdNiO$_2$~\cite{PhysRevLett.125.027001, PhysRevLett.125.147003}.
This fact suggests that the SC state in this material might be in a strong interplay with the CDW fluctuations.
A similar conclusion was also reported in a recent experimental work~\cite{ji2022rotational}, where the rotational symmetry breaking observed in superconducting Nd$_{0.8}$Sr$_{0.2}$NiO$_{2}$ films was associated with the charge ordering.

\section{Results}

\paragraph{\bf Model.} 
In order to describe explicit many-body effects in NdNiO$_2$ we use a minimal 3-orbital ${\{d_{z^2}, d_{x^2-y^2}, \text{SD}\}}$ model that accounts for an almost occupied Ni-$d_{z^2}$ orbital, a half-filled Ni-$d_{x^2-y^2}$ orbital, and a nearly empty self-doping (SD) band. To this end, the 3-orbital Wannier Hamiltonian derived in Ref.~\cite{PhysRevX.10.041002} from DFT calculations for NdNiO$_2$ is taken and supplemented with local interactions.
The resulting three-orbital Hubbard-Holstein-Kanamori Hamiltonian reads: 
\begin{align}
H = &\sum_{jj',\sigma,ll'} t^{ll'}_{jj'} c^{\dagger}_{j\sigma{}l} c^{\phantom{\dagger}}_{j'\sigma{}l'} 
+ \frac12 \sum_{\substack{j, \{l\},\{\sigma\}}} U^{\phantom{\dagger}}_{l_1 l_2 l_3 l_4} c^{*}_{j\sigma l_1} c^{*}_{j\sigma' l_2} c^{\phantom{*}}_{j\sigma' l_4} c^{\phantom{*}}_{j\sigma l_3} \notag\\
&+\omega_{\rm ph}\sum_{j}b^{\dagger}_{j}b^{\phantom{\dagger}}_{j} 
+ \lambda\sum_{j} n^{\phantom{\dagger}}_{j}\left(b^{\phantom{\dagger}}_{j}+b^{\dagger}_{j}\right),
\label{eq:H_latt}
\end{align}
where $c^{(\dagger)}_{j\sigma{}l}$ operator describes annihilation (creation) of an electron on the site $j$ on the band $l$ with the spin projection ${\sigma=\{\uparrow,\downarrow\}}$.
The dispersion of the electrons is defined by the hopping amplitudes $t^{ll'}_{jj'}$ that are obtained from \emph{ab-initio} calculations.
The on-site interaction $U_{l_1 l_2 l_3 l_4}$ for ${l_i\in\{d_{z^2}, d_{x^2-y^2}\}}$ bands is taken in the Kanamori form that includes the intraorbital ${U_{llll}=U}$, interorbital ${U_{ll'll'}=U'=U-2J}$, spin flip ${U_{ll'l'l}=J}$, and pair hopping ${U_{lll'l'}=J}$ terms. 
The intraorbital term is set to ${U=7}$\,eV and the Hund's rule coupling to ${J=1}$\,eV~\cite{PhysRevX.10.041002}.
The SD band is considered uncorrelated, so ${U_{l_1 l_2 l_3 l_4}=0}$ if any index $l_i$ belongs to SD. 

The second line in Eq.~\eqref{eq:H_latt} describes the effect of phonons. 
The operator $b^{(\dagger)}_{j}$ annihilates (creates) a phonon on the site $j$. The phonon operators are coupled to the electronic density operator defined as ${n^{\phantom{\dagger}}_{j}=\sum_{j,\sigma,l}c^{\dagger}_{j\sigma{}l} c^{\phantom{\dagger}}_{j\sigma{}l}}$.
The phonon frequency ${\omega_{\rm ph}=24.4}$\,meV (283\,K) and the electron-phonon coupling ${\lambda=0.22}$ are approximated with a local phonon model following Ref.~\cite{PhysRevB.100.205138}.
For convenience, we integrate out phonon operators, which results in an effective local frequency-dependent attractive interaction
${U^{\rm ph}_{ll'll'}(\omega) = -2\lambda^2\frac{\omega_{\rm ph}}{\omega^2_{\rm ph}-\omega^2}}$ between the electronic densities~\cite{PhysRevB.52.4806, PhysRevLett.94.026401, PhysRevLett.99.146404, stepanov2021coexisting}.

\begin{figure}[t!]
\includegraphics[width=1\linewidth]{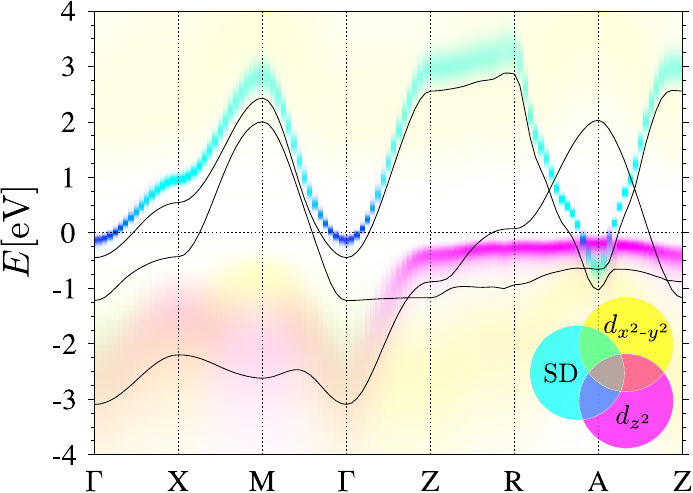}
\caption{The momentum-resolved electronic spectral function. The result is obtained at half filling for ${\mu_{xy}}$ using LDA (solid black lines) and DMFT (see the inset for the color code) for the $d_{z^2}$, $d_{x^2-y^2}$, and SD bands along the high-symmetry path in the BZ.
\label{fig:LDA_DMFT}}
\end{figure}

Strong electronic correlations in NdNiO$_2$ manifest themselves in the formation of Hubbard bands and in the orbital-selective Mott behavior of the material~\cite{PhysRevX.10.041002}. 
Taking these effects into account requires using DMFT that accurately describes local electronic correlations.  
Addressing many-body instabilities related to spatial collective electronic fluctuations, e.g. the formation of the CDW ordering, requires using diagrammatic extensions of DMFT~\cite{RevModPhys.90.025003, Lyakhova_review} that can efficiently treat momentum-dependent electronic correlations.
We use the dual triply irreducible local expansion (\mbox{D-TRILEX}) method~\cite{PhysRevB.100.205115, PhysRevB.103.245123, 10.21468/SciPostPhys.13.2.036}, which consistently accounts for the effect of local correlations and spatial collective electronic fluctuations of an arbitrary range within a broad set of model parameters~\cite{stepanov2021coexisting, vandelli2024doping, Maria, Zeros}.
In this method, local electronic correlations are taken into account exactly by solving the DMFT impurity problem using continuous time quantum Monte Carlo solvers~\cite{PhysRevB.72.035122, PhysRevLett.97.076405, PhysRevLett.104.146401, RevModPhys.83.349}. 
In this work, the impurity problem is solved using the \textsc{w2dynamics} package~\cite{WALLERBERGER2019388}. 
The spatial collective electronic fluctuations are treated approximately by considering the leading (particle-hole ladder-like) diagrammatic contributions to the self-energy and the polarization operator~\cite{PhysRevB.100.205115, PhysRevB.103.245123, 10.21468/SciPostPhys.13.2.036}.
In this way, the method provides a self-consistent many-body solution to the problem at the level of both, single- (electronic Green's function) and two-particle (charge, spin, and orbital susceptibilities) response functions with full momentum and frequency dependence.
The advantage of \mbox{D-TRILEX} is that it has a much simpler diagrammatic structure compared to other extensions of DMFT, enabling tractable numerical calculations in the multi-orbital framework~\cite{PhysRevLett.127.207205, PhysRevResearch.5.L022016, PhysRevLett.129.096404, stepanov2023orbitalselective}.
In particular, the \mbox{D-TRILEX} method does not involve the four-point vertex function, the calculation~\cite{RevModPhys.83.349, WALLERBERGER2019388, 10.21468/SciPostPhys.8.1.012} and utilization of which in the diagrammatic expansion~\cite{PhysRevB.95.115107, doi:10.7566/JPSJ.87.041004, PhysRevB.103.035120} significantly complicates numerical calculations in the multi-orbital case. 
Despite a rather simple structure, \mbox{D-TRILEX} operates on the same level of accuracy as much more advanced diagrammatic approaches~\cite{PhysRevB.103.245123, 10.21468/SciPostPhys.13.2.036, PhD_Vandelli}, which is achieved by considering important local three-point (Hedin~\cite{GW1}) vertex corrections in the self-energy and the polarization operator.
The main drawback of the current method implementation is its restriction to the non-symmetry-broken phase.
This limitation is shared with the majority of other diagrammatic extensions of DMFT, although there have been several attempts to perform diagrammatic calculations in the symmetry-broken phases (see, e.g., Refs.~\cite{PhysRevLett.121.037204, PhysRevLett.122.127601, PhysRevB.104.085120, PhysRevB.107.245104, delre2024twoparticle}). 
It means, that \mbox{D-TRILEX} can capture the formation of the ordered state by the divergence of the corresponding susceptibility (the momentum ${\bf q}$, at which the susceptibility diverges, corresponds to the wave vector of the ordering), but cannot perform calculations inside the ordered phases.
This limitation sometimes restricts \mbox{D-TRILEX} calculations to a rather high-temperature regime if the system reveals magnetic instability already at high temperatures, as happens, e.g., in the considered case.
Details on the many-body calculations are provided in the Methods section.

\paragraph{\bf Formation of the CDW at half filling.}
The half-filled case of ${n_{\rm total}=3}$ electrons per three ${\{d_{z^2}, d_{x^2-y^2}, \text{SD}\}}$ bands corresponds to NdNiO$_2$ at stoichiometry.
DFT in local-density approximation (LDA) (i.e. the formal ${U=0}$ case) predicts an exotic band structure for the system with the following occupation of the bands: ${\{n_{z^2}, n_{x^2-y^2}, n_{\rm SD}\} = \{1.84, 0.93, 0.23\}}$~\cite{PhysRevX.10.041002}. 
The electronic band structure of LDA is shown in Fig.~\ref{fig:LDA_DMFT} in solid black lines. 
The result is calculated along the high-symmetry path in the Brillouin Zone (BZ) that consists of the ${\Gamma = (0,0,0)}$, ${\text{X}=(\pi,0,0)}$, ${\text{M}=(\pi,\pi,0)}$, ${\text{Z}=(0,0,\pi)}$, ${\text{R}=(\pi,0,\pi)}$, and ${\text{A}=(\pi,\pi,\pi)}$ points.
According to LDA~\cite{PhysRevX.10.041002} the $d_{z^2}$ band is almost fully filled and features a flat part that, however, lies near the Fermi energy only at ${k_{z}=\pi}$ momentum.
In addition, the $d_{z^2}$ band has an electron pocket around the $\Gamma$ point that is hybridized with a slightly unoccupied SD band.
The $d_{x^2-y^2}$ orbital appears to be metallic and nearly half-filled.

Local electronic correlations accounted for in the charge self-consistent combination~\cite{PhysRevB.86.155121} of DFT, self-interaction correction (SIC), and DMFT (DFT+sicDMFT framework~\cite{PhysRevB.100.115125}) change the occupation of bands to $\{1.83, 1.00, 0.17\}$ and consequently modify the electronic spectral function~\cite{PhysRevX.10.041002}.
The latter is shown in Fig.~\ref{fig:LDA_DMFT} (see the inset for the color code).
One finds, that considering local correlations results in the shift of the flat part of the $d_{z^2}$ band (high-intensity weight plotted in magenta) closer to the Fermi energy and also makes the $d_{x^2-y^2}$ orbital Mott insulating (low-intensity weight at large energies plotted in yellow).
Nevertheless, the $d_{z^2}$ (magenta) and SD (cyan) bands remain metallic, which makes the NdNiO$_2$ an orbital-selective Mott insulator~\cite{PhysRevX.10.041002}.

To account for nonlocal correlation effects beyond DMFT, we perform \mbox{D-TRILEX} calculations for the model Hamiltonian~\eqref{eq:H_latt} that describes the low-energy part of the electronic spectrum.
In this model, only two of three orbitals are considered correlated, so introducing electronic interactions will shift the position of correlated $d_{z^2}$ and $d_{x^2-y^2}$ bands with respect to the uncorrelated SD one.
In order to reproduce the correct position of the bands, one has to compensate this shift by introducing the double-counting (DC) correction, i.e. the constant shift $\mu^{\rm DC}$, for the $d_{z^2}$ and $d_{x^2-y^2}$ orbitals (see, e.g., Ref.~\cite{PhysRevX.10.041002}).
In the DFT+sicDMFT framework the DC correction is usually determined from the self-consistent calculation of the Hartree-Fock-like shift, which is related to the occupation of orbitals.
In \mbox{D-TRILEX}, a self-consistent determination of $\mu^{\rm DC}$ would require solving the DMFT impurity problem for the single- and two-particle quantities and performing the \mbox{D-TRILEX} diagrammatic calculation at every iteration of a self-consistent cycle.
This procedure is enormously time-consuming and is unfeasible in practice.
We should also note that the value of the DC correction determined in this way is not exact, as it is related to the Hartree-Fock-like approximation. 
In this work we chose another root and perform a scan over a certain range of DC corrections following Ref.~\cite{KAROLAK201011}.
This procedure allows us to investigate sensitivity of correlation effects to $\mu^{\rm DC}$ in the unique case when one of the correlated orbitals ($d_{z^2}$) features the flat band near the Fermi level.
The resulting occupation of the bands obtained for different values of the DC correction is compared to the ones of the DFT+sicDMFT scheme to determine the most consistent $\mu^{\rm DC}$.
To identify the range of values for $\mu^{\rm DC}$ we calculate the Hartree-Fock shift induced by electronic interactions for the $d_{z^2}$ and $d_{x^2-y^2}$ orbitals.
This shift can be obtained from the form of the interaction part of the Hamiltonian~\eqref{eq:H_latt} and reads:
\begin{align}
\mu^{\rm DC}_{z^2} &= \frac{U}{2} n_{z^2} + \frac{2U' - J}{2} n_{x^2-y^2}, \\
\mu^{\rm DC}_{x^2-y^2} &= \frac{U}{2} n_{x^2-y^2} + \frac{2U' - J}{2} n_{z^2}.
\end{align}
Substituting the DFT+sicDMFT values for the occupations gives ${\mu^{\rm DC}_{z^2} = 10.9\equiv\mu_z}$ and ${\mu^{\rm DC}_{x^2-y^2} = 11.7\equiv\mu_{xy}}$, which sets the lower and upper bounds for $\mu^{\rm DC}$, respectively.

\begin{figure*}[t!]
\includegraphics[width=1\linewidth]{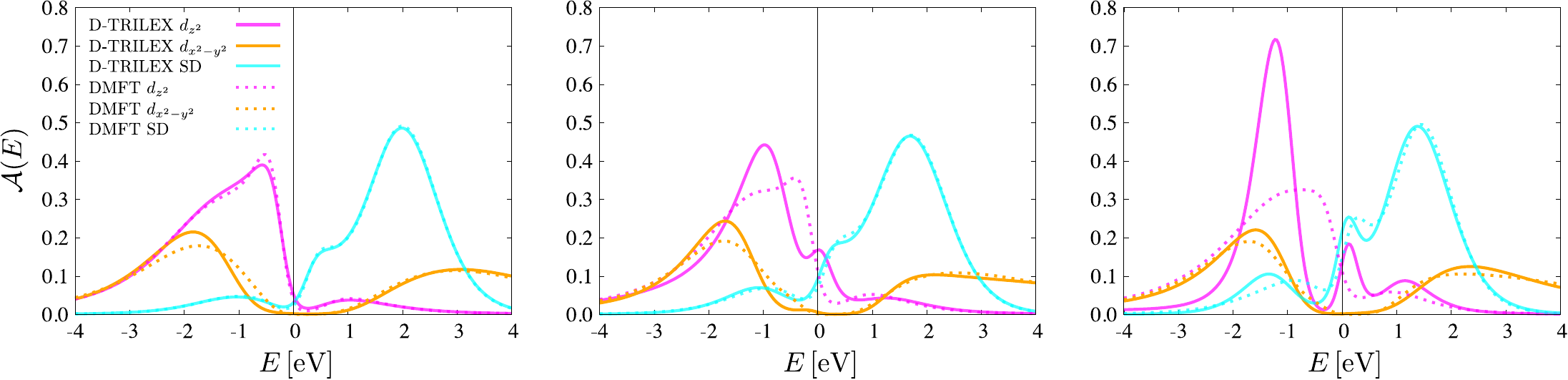}
\caption{Band-resolved local electronic spectral functions. Results are obtained for $d_{z^2}$ (magenta), $d_{x^2-y^2}$ (orange), and SD (cyan) bands for $\mu_{xy}$ (left panel), ${\mu^{\rm DC}=11.2}$ (middle panel), and ${\mu_{z}}$ (right panel). The \mbox{D-TRILEX} and DMFT results are shown in solid and dashed lines, respectively.
The system demonstrates the orbital-selective Mott behavior: The $d_{x^2-y^2}$ is half-filled and displays two Hubbard bands. The $d_{z^2}$ and SD bands remain metallic for all considered $\mu^{\rm DC}$. 
Decreasing $\mu^{\rm DC}$ results in a stronger attraction between the $d_{z^2}$ and SD bands leading to the development of sharp peaks in each spectral function that merge in the vicinity of the Fermi energy at ${\mu_{z}}$.
\label{fig:DOS}}
\end{figure*}

For the identified range of DC values we first perform calculations in the absence of the electron-phonon coupling for the fixed total occupation of ${n_{\rm total}=3}$ electrons per three bands.
We begin by solving the DMFT impurity problem, which is done separately for each considered value of $\mu^{\rm DC}$.
After that, the exact impurity Green's function, susceptibilities and vertex functions are used to construct the \mbox{D-TRILEX} diagrammatic expansion~\cite{PhysRevB.100.205115, PhysRevB.103.245123, 10.21468/SciPostPhys.13.2.036}.
Within this work all calculations are done for a fixed temperature ${T=0.1}$\,eV.
At this temperature the system lies in the paramagnetic phase close to the magnetic instability.
In order to estimate the strength of the charge and spin fluctuations, we use the leading eigenvalue (LEV) of the Bethe-Salpeter equation (BSE) for the charge and spin susceptibility~\cite{10.21468/SciPostPhys.13.2.036}. 
The spin LEV at the chosen temperature is ${\text{LEV}\simeq0.75}$ as shown in Table~\ref{tab:occ}. 
Therefore, in this temperature regime the spin fluctuations are already strong, as ${\text{LEV}=1}$ would indicate the divergence of the susceptibility upon the transition to the ordered state.
The corresponding eigenvector ${{\bf q} = \text{M}}$ shows that the leading instability in the spin channel corresponds to an in-plane (C-type) antiferromagnetic (AFM) ordering, which is consistent with experimental~\cite{Cui_2021} observations and previous theoretical predictions~\cite{PhysRevX.10.021061, gu2020substantial, PhysRevMaterials.5.044803, PhysRevB.105.205131}. 

\begin{table}[b!]
\caption{The occupation of bands $n_l$, the LEV of the BSE for the charge and spin susceptibilities and the corresponding eigenvectors ${\bf q}$. Results are obtained for different values of $\mu^{\rm DC}$. 
All calculations, except for the last row, are performed in the absence of the electron-phonon (e-ph) coupling.}
\begin{tabular}{|c|c|c|c|c|c|}
    \hline
$\mu^{\rm DC}$ & $n_{z^2}$ & $n_{x^2-y^2}$ & $n_{\rm SD}$ & ~charge LEV~({\bf q})~ & ~spin LEV~({\bf q})~ \\
    \hline
~~11.7~~ & ~1.83~ & ~1.00~ & ~0.17~ & ~0.02~(M)~ & ~0.72~(M)~  \\
~~11.5~~ & ~1.80~ & ~1.00~ & ~0.20~ & ~0.16~(M)~ & ~0.75~(M)~  \\
~~11.4~~ & ~1.78~ & ~1.00~ & ~0.22~ & ~0.34~(X)~ & ~0.75~(M)~  \\
~~11.2~~ & ~1.73~ & ~1.00~ & ~0.27~ & ~0.64~(X)~ & ~0.77~(M)~  \\
~~11.0~~ & ~1.70~ & ~1.00~ & ~0.30~ & ~0.83~(X)~ & ~0.76~(M)~  \\
~~10.9~~ & ~1.65~ & ~1.00~ & ~0.35~ & ~0.95~(X)~ & ~0.74~(M)~  \\
\hline
~10.9 (e-ph)~ & ~1.62~ & ~1.00~ & ~0.38~ & ~0.95~(X)~ & ~0.74~(M)~  \\
\hline
\end{tabular} 
\label{tab:occ}
\end{table}

The occupation of bands obtained in \mbox{D-TRILEX} for different $\mu^{\rm DC}$ are specified in Table~\ref{tab:occ}.
These results suggest that ${\mu_{xy}}$ is the most consistent value of the DC correction, as it identically reproduces the occupation of bands found in DFT+sicDMFT~\cite{PhysRevX.10.041002}.
In agreement with the DFT+sicDMFT calculations we find that taking into account correlation effects makes the $d_{x^2-y^2}$ band half filled for every considered value of $\mu^{\rm DC}$: ${n_{x^2-y^2}=1.00}$ in \mbox{D-TRILEX} instead of ${n_{x^2-y^2}=0.93}$ in LDA.
This fact explains why the change in $\mu^{\rm DC}$ makes almost no influence on the strength (spin LEV, Table~\ref{tab:occ}) of spin fluctuations.
We note, that for the largest considered value of the DC correction ${\mu_{xy}}$ the occupation of bands is redistributed by reducing $n_{\rm SD}$, while the filling of the $d_{z^2}$ band remains nearly unchanged compared to the LDA value.
We also note, that for ${\mu_{xy}}$ the occupation of bands predicted by \mbox{D-TRILEX} coincides with one obtained within the DFT+sicDMFT scheme~\cite{PhysRevX.10.041002}.
On the contrary, for ${\mu^{\rm DC}\simeq11.4}$ the situation is reversed, and now the half filling of the $d_{x^2-y^2}$ band is achieved by reducing the occupation of the ${d_{z^2}}$ band, while the occupation of the SD band is similar to the LDA one.
Reducing the value of the DC correction to $\mu_{z}$ redistributes the occupation of bands in such a way, that the filling of the SD band becomes substantially larger than the one obtained in LDA and in DFT+sicDMFT.

In contrast to the nearly unchanged spin LEV, the charge LEV increases substantially with decreasing ${\mu^{\rm DC}}$.
Thus, Table~\ref{tab:occ} shows that for the most consistent DC correction ${\mu_{xy}}$ the charge fluctuations are basically absent in the system (${\text{LEV}=0.02}$).
Reducing $\mu^{\rm DC}$ to ${\mu_z}$ drives the system in the region close to the CDW instability (${\text{LEV}=0.95}$) that is characterised by the ordering vector ${{\bf q}_{\rm CDW} = \text{X}}$.
We find, that the CDW ordering vector remains unchanged for all values of ${\mu^{\rm DC}}$ at which the charge fluctuations are well developed (${\text{LEV}\geq0.34}$). 
Remarkably, we find that the formation of the CDW is unrelated to the usual mechanisms behind the charge instability, namely the electron-phonon coupling and long-range Coulomb interaction that were not taken into account in these calculations.
As we show below, considering phonon degrees of freedom does not affect the formation of the CDW ordering.
In turn, the realistic value of the nonlocal Coulomb interaction in NbNiO$_2$ is too small ($V\simeq{}U/12$~\cite{PhysRevB.100.205138}) to make any influence on the charge instability.

\begin{figure*}[t!]
\includegraphics[width=1\linewidth]{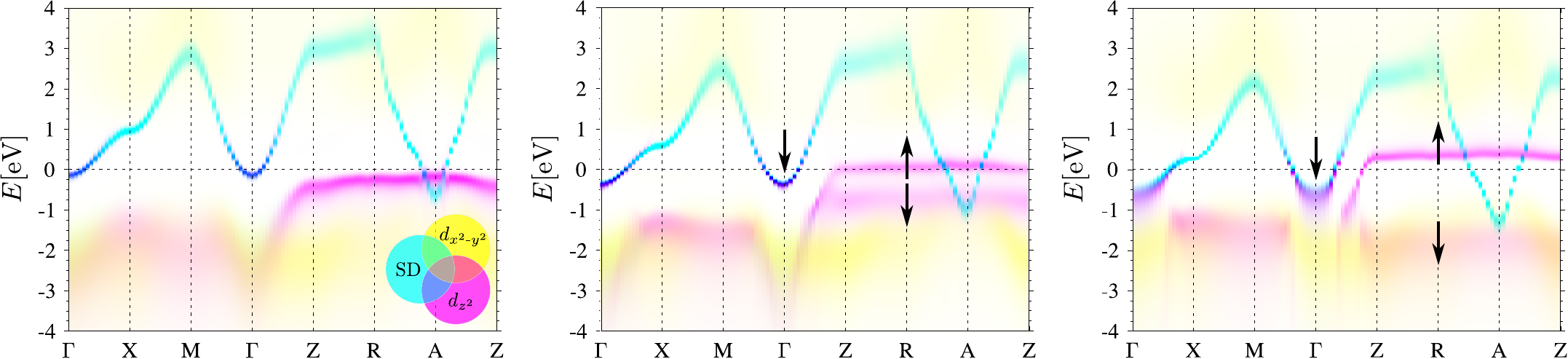}
\caption{Momentum-resolved electronic spectral functions. Results are obtained for the $d_{z^2}$, $d_{x^2-y^2}$, and SD bands (see the inset in the right panel for the color code) for ${\mu_{xy}}$ (left panel), ${\mu^{\rm DC}=11.2}$ (middle panel), and ${\mu_z}$ (right panel) along the high-symmetry path in the BZ.
The low-intensity part of the spectrum at high energies plotted in yellow corresponds to Hubbard bands of the half-filled $d_{x^2-y^2}$ orbital.
The high-intensity part of the spectrum around the Fermi energy plotted in magenta corresponds to the $d_{z^2}$ orbital that displays a flat band feature at ${k_{z}=\pi}$. 
The black arrows depict the shift of the $\Gamma$ point and of the flat part of the $d_{z^2}$ band upon decreasing $\mu^{\rm DC}$. 
\label{fig:A}}
\end{figure*}

To identify the source of the CDW instability let us first look at the local electronic spectral function ${\cal A}(E)$ shown in Fig.~\ref{fig:DOS} for the three different values of $\mu^{\rm DC}$. 
We find that for the most consistent ${\mu_{xy}}$ the spectral function of \mbox{D-TRILEX} (solid lines) is very similar to the one of DMFT (dashed lines).
Both methods predict a Mott insulating behavior for the half-filled $d_{x^2-y^2}$ band, while the $d_{z^2}$ and SD bands remain metallic for all values of ${\mu^{\rm DC}}$.
Therefore, the \mbox{D-TRILEX} approach also finds an orbital-selective Mott insulating behavior for NdNiO$_2$ in agreement with the previous DMFT calculation~\cite{PhysRevX.10.041002}.
It is important to point out that this state is not destroyed by the magnetic fluctuations~\cite{PhysRevLett.129.096404}, because the metallic $d_{z^2}$ and SD bands have a non-integer filling (see Table~\ref{tab:occ}), which suppresses the strength of the magnetic fluctuations.
We observe, that decreasing $\mu^{\rm DC}$ makes the DMFT result qualitatively unchanged, but the form of the local spectral function predicted by \mbox{D-TRILEX} changes substantially.
First, we note that in \mbox{D-TRILEX} the Hubbard $d_{z^2}$ bands, which for ${\mu_{xy}}$ appear at ${E\simeq-2}$\,eV and ${E\simeq3}$\,eV, move closer to the Fermi energy $E_{F}$.
Second, reducing the value of ${\mu^{\rm DC}}$ increases the attraction between the $d_{z^2}$ and SD bands. 
This fact is evident from the development of peaks in the corresponding (magenta and cyan) \mbox{D-TRILEX} spectral functions near the Fermi energy $E_{F}$ (middle panel) that eventually merge at the same energy (right panel) when the system approaches the CDW transition point.
This attraction seems to be realized through the spatial collective electronic fluctuations as it is not captured by DMFT.

To investigate the effect of spatial electronic correlations in more detail, let us look at the the momentum-resolved spectral function shown in Fig.~\ref{fig:A} for the $d_{z^2}$ (magenta), $d_{x^2-y^2}$ (yellow), and SD (cyan) bands. 
The result is obtained using \mbox{D-TRILEX} for $\mu_{xy}$ (left panel), ${\mu^{\rm DC}=11.2}$ (middle panel), and $\mu_{z}$ (right panel).
The dispersiveless low-intensity ``yellow'' weight at small and large energies corresponds to the Hubbard $d_{x^2-y^2}$ bands.
The dispersive high-intensity ``magenta'' weight near the Fermi energy is the metallic $d_{z^2}$ band that displays a flat feature at ${k_{z}=\pi}$. 
We point out, that for all considered values of $\mu^{\rm DC}$ the $d_{z^2}$ band reveals a momentum-dependent hybridization with the other bands.
The hybridization of the $d_{z^2}$ and SD bands around the $\Gamma$ point results in the formation of the electron pocket.
At ${k_{z}=0}$ the $d_{z^2}$ band hybridizes with the lower Hubbard $d_{x^2-y^2}$ band that lies at ${E\simeq-2}$\,eV.

\begin{figure}[b!]
\includegraphics[width=0.48\linewidth]{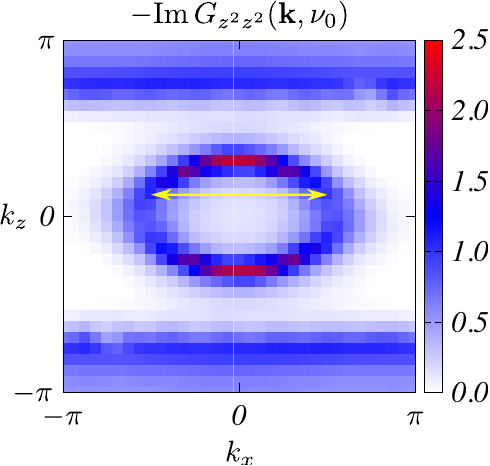} ~~
\includegraphics[width=0.48\linewidth]{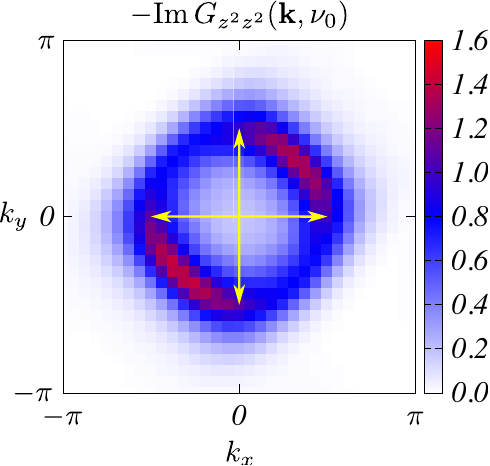}
\caption{The Fermi surface. The 2D cuts of the Fermi surface in the $(k_x,0,k_z)$ (left panel) and $(k_x,k_y,\pi/8)$ (right panel) calculated near the CDW instability for ${\mu_z}$. The FS is approximated by the imaginary part of the lattice Green's function taken at the zeroth Matsubara frequency ${-\text{Im}\,G_{z^2z^2}({\bf k},\nu_0)}$. The CDW ordering vectors ${{\bf q}_{\rm CDW}}$ connecting FS are shown in yellow arrows. 
\label{fig:FS}}
\end{figure}

\begin{figure*}[t!]
\includegraphics[width=1\linewidth]{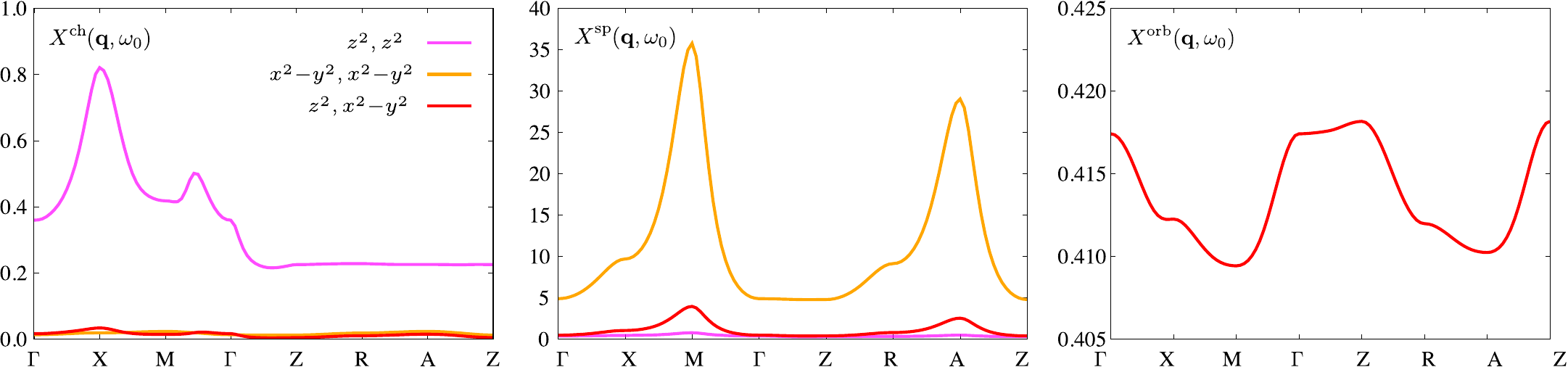}\
\caption{Static susceptibilities. The absolute value of the band-resolved static charge (left panel), spin (middle panel) and orbital (right panel) susceptibility $X^{\rm ch/sp}_{ll'}({\bf q})$ calculated for ${l,l'\in\{d_{z^2},d_{x^2-y^2}\}}$ bands along the high-symmetry path in the BZ. The leading contributions to the charge and spin susceptibilities originate respectively from the intraband collective electronic fluctuations in the $d_{z^2}$ and $d_{x^2-y^2}$ bands. The orbital fluctuations between the $d_{z^2}$ and $d_{x^2-y^2}$ bands are rather small and predominantly local.
The result is obtained in the proximity to the CDW instability achieved for the DC correction $\mu_{z}$ .
\label{fig:X}}
\end{figure*}

We find that tuning $\mu^{\rm DC}$ leads to a strong momentum-dependent renormalization of the electronic spectral function, and, in particular, of the $d_{z^2}$ band.
This effect cannot be captured on the basis of local theories and requires an advanced combination of the band structure theory with the momentum-dependent many-body approach.
The most striking change concerns the flat part of the $d_{z^2}$ band.
At the largest DC correction $\mu_{xy}$ this part lies below $E_{F}$ (left panel in Fig.~\ref{fig:A}).
Upon decreasing ${\mu^{\rm DC}}$ the flat band splits into two parts that move in the opposite directions in energy as depicted by the black arrows in Fig.~\ref{fig:A}. 
One part moves toward the Fermi energy and at ${\mu^{\rm DC}=11.2}$ appears at $E_{F}$ (middle panel).
The other part moves toward lower energies and at $\mu_{z}$ hybridizes with the lower Hubbard $d_{x^2-y^2}$ band (right panel). 
Remarkably, at $\mu_{z}$, when the system approaches the CDW transition point, the upper part of the flat band moves above the Fermi energy and becomes unoccupied.
From this fact one can conclude that the CDW instability in NdNiO$_2$ is not related to the flat band feature.
Remarkably, we find that in the vicinity of the CDW phase transition the position in energy of the flat part of the $d_{z^2}$ band coincides with the position of the van Hove singularity (vHS) of the SD band located at the X point.
This fact was observed in the local spectral function as the attraction between the $d_{z^2}$ and SD bands manifested itself with two (magenta and cyan) matching peaks near $E_F$ (right panel in Fig.~\ref{fig:DOS}).

Another drastic change in the spectral function occurs in the vicinity of the $\Gamma$ point, where the hybridized $d_{z^2}$ and the SD bands form an electron pocket. 
We note that this pocket contains a relatively large spectral weight.
Upon reducing $\mu^{\rm DC}$ the $\Gamma$ point moves to lower energies causing an enhancement of the electron pocket in order to compensate the shift of the upper part of the flat band in the opposite direction.
These observations suggest that the CDW instability originates from the nesting of the Fermi surface (FS) related to the electron pocket of the $d_{z^2}$ band, as the SD band is uncorrelated.
Remarkably, at $\mu_{z}$, when the system is in the vicinity of the CDW phase transition, the $\Gamma$ point moves to such low energies that it leads to a hybridization of the $d_{z^2}$ and SD bands with the lower Hubbard $d_{x^2-y^2}$ band at this k-point (right panel in Fig.~\ref{fig:A}).

We approximate the FS by the imaginary part of the lattice Green's function (${-\text{Im}\,G({\bf k},\nu_n)}$) taken at the zeroth fermionic Matsubara frequency $\nu_0$.  
The CDW ordering vector ${{\bf q}_{\rm CDW} = \text{X}}$ suggests that the two-particle scattering between the FS points occurs at a constant momenta $k_z$ and $k_x$ (or $k_y$). 
For this reason, in Fig.~\ref{fig:FS} we plot the 2D cuts of the FS in the $(k_x,0,k_z)$ (left panel) and $(k_x,k_y,\pi/8)$ (right panel) obtained in the vicinity of the CDW ordering for ${\mu_{z}}$.
Plotting the ${\bf q}_{\rm CDW}$ vectors explicitly (yellow arrows) suggests that the electronic scattering that leads to the formation of the CDW ordering occurs for ${k_z\simeq\pi/8}$ and ${k_x\,(k_y)=0}$. 

The conclusion that the CDW instability originates from the electronic scattering within the $d_{z^2}$ band is also confirmed by the form of the charge susceptibility. 
Fig.~\ref{fig:X} shows the absolute value of the static (${\omega_0}$) \mbox{D-TRILEX} charge (left panel), spin (bottom panel) and orbital (right panel) susceptibility~\cite{10.21468/SciPostPhys.13.2.036} $X^{\rm ch/sp}_{ll'}({\bf q},\omega_n)$ obtained for the correlated ${l,l'\in\{d_{z^2},d_{x^2-y^2}\}}$ bands along the high-symmetry path in the BZ. 
We find, that the dominant contribution to the charge susceptibility originates from the intraband $d_{z^2}$ component that reveals a peak at exactly the ${\bf q}_{\rm CDW}$ wave vector.
The charge susceptibility also displays a subleading mode corresponding to the wave vector ${{\bf q} \simeq (0.3\pi, 0.3\pi, 0)}$.
We note, that the previous DMFT calculation performed for the distorted crystal structure corresponding to the CDW phases obtained in DFT calculations predicted the ${{\bf q} = (\pi/3, \pi/3, 0)}$ wave vector for the CDW modulation~\cite{PhysRevB.106.165110}.
On the contrary, the leading contribution to the spin susceptibility comes from the intraband $d_{x^2-y^2}$ component.
The highest peak in the spin susceptibility at the ${{\bf q} = \text{M}}$ point indicates that the leading magnetic instability corresponds to the in-plane AFM (C-AFM) ordering.
This fact is also evident from the eigenvector ${{\bf q} = \text{M}}$ corresponding to the LEV of the spin fluctuations (Table~\ref{tab:occ}), as has been discussed above. 
However, a relatively high peak at the ${{\bf q} = \text{A}}$ point indicates that the G-type AFM fluctuations in the system are also rather strong. 
The competition between the C- and G-types of AFM ordering has also been predicted for doped NdNiO$_2$ in Ref.~\cite{LEONOV2021160888} based on the DMFT calculations. However, the G-type of AFM ordering was found to be dominant at stoichiometry.
Finally, by looking at the right panel of Fig.~\ref{fig:X}, one finds that the orbital fluctuations between the $d_{z^2}$ and $d_{x^2-y^2}$ bands are negligibly small and mainly local (constant in momentum space).

\begin{figure}[t!]
\includegraphics[width=1\linewidth]{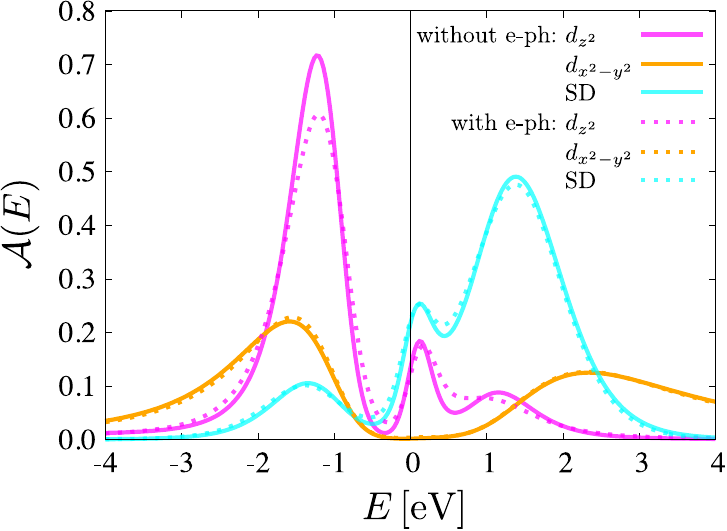}
\caption{The band-resolved local electronic spectral function. Results are obtained for $d_{z^2}$ (magenta), $d_{x^2-y^2}$ (orange), and SD (cyan) bands without (solid lines) and with (dashed line) taking into account the electron-phonon (e-ph) coupling for ${\mu_z}$.
\label{fig:DOS_e-ph}}
\end{figure}

In order to illustrate the influence of the phonon degrees of freedom on the formation of the CDW ordering we repeat the same calculations in the presence of the electron-phonon coupling.
We find that, besides a small difference in the occupation of the $d_{z^2}$ and SD bands (bottom row of Table~\ref{tab:occ}), considering phonons does not affect the obtained results.
Indeed, the strength and the ordering vectors of the charge and spin fluctuations remain the same as in the absence of the electron-phonon coupling (bottom row of Table~\ref{tab:occ}), and the electronic spectral function is barely changed (Fig.~\ref{fig:DOS_e-ph}). 
The only noticeable difference is the fact that the half filling in these two cases is realized for different values of the chemical potential.
To be precise, taking into account the electron phonon coupling results in the increase of the chemical potential by ${\delta\mu\simeq0.066}$\,eV.
We note, however, that the Holstein model for phonon degrees of freedom may lead to an underestimate of the phonon contribution to the CDW instability, since in this case the phonon-mediated electronic interaction $U^{\rm ph}_{ll'll'}(\omega)$ competes with the large intraorbital $U$ and interorbital $U'$ Coulomb interactions.
Considering other types of the electron-phonon coupling, e.g. a Jahn-Teller or a momentum-dependent ones, may lead to a more noticeable effect~\cite{PhysRevB.107.085131}.

In both cases (with and without accounting for the electron-phonon coupling) the large leading eigenvalue of the charge fluctuations ${\text{LEV}=0.95}$ is found for the same DC correction ${\mu_z}$ (Table~\ref{tab:occ}).
However, at this value of ${\mu^{\rm DC}}$ the occupation of the SD band is already approximately two times larger and the occupation of the $d_{z^2}$ band is substantially lower than the ones obtained within the DFT+sicDMFT framework~\cite{PhysRevX.10.041002}.
In addition, the local spectral function calculated for ${\mu_{xy}}$ (left panel in Fig.~\ref{fig:A}) much better reproduces the experimentally observed photoemission spectrum~\cite{doi:10.1073/pnas.2007683118, CHEN20221806} than the one obtained for ${\mu_z}$ (right panel in Fig.~\ref{fig:A}). 
Indeed, a dominant first peak at ${E\simeq-2}$\,eV seen in the experiments agrees well with the position of the lower Hubbard $d_{x^2-y^2}$ band obtained for ${\mu_{xy}}$.
For smaller $\mu^{\rm DC}$ the Hubbard bands are shifted closer to the Fermi level.
Furthermore, the experiments also do not observe sharp peaks due to a flat part of the $d_{z^2}$ band appearing in the vicinity of $E_{F}$ at stoichiometry.
Note, that in DFT+sicDMFT such shift of the flat band only takes place for a sizable doping~\cite{PhysRevX.10.041002}.
Therefore, the choice of ${\mu_z}$ does not correspond to the pristine NdNiO$_2$.
On the other hand, in the case of ${\mu_{xy}}$, which reproduces the DFT+sicDMFT occupation of bands, the charge fluctuations are absent in the system, as discussed above.
Therefore, one can conclude that the CDW ordering cannot be found in NdNiO$_2$ at stoichiometry, which is in line with the results of recent experiments~\cite{https://doi.org/10.1002/smll.202304872, parzyck2023, hayashida2024investigation}.

\paragraph{\bf CDW upon hole doping.}
According to the results obtained at half filling the CDW instability in NdNiO$_2$ originates from the intraband scattering of electrons within the $d_{z^2}$ electron pocket and arises upon reducing the occupation of this band.
The latter might be, in principle, achieved in a more physical way, i.e. upon hole doping the system. 
In order to understand if this results in the formation of the CDW ordering, we stick to the most consistent DC correction ${\mu_{xy}}$ and perform the \mbox{D-TRILEX} calculations for different levels of the hole doping. 
Similarly to the half-filled case considered above, we find that considering the electron-phonon coupling does not affect the physical behavior of the system. 
For this reason, in this section, we present only the results calculated in the presence of the electron-phonon coupling.

\begin{table}[t!]
\caption{The occupation of bands $n_l$, the LEV of the BSE for the charge and spin susceptibilities and the corresponding eigenvectors ${\bf q}$. Results are obtained for ${\mu_{xy}}$ in the presence of the electron-phonon coupling for different levels of the hole doping. 
}
\begin{tabular}{|c|c|c|c|c|c|}
    \hline
~$n_{\rm total}$~ & $n_{z^2}$ & $n_{x^2-y^2}$ & $n_{\rm SD}$ & ~charge LEV~({\bf q})~ & ~spin LEV~({\bf q})~ \\
    \hline
~2.96~ & ~1.78~ & ~1.00~ & ~0.18~ & ~0.49~(X)~ & ~0.74~(M)~  \\
~2.92~ & ~1.75~ & ~1.00~ & ~0.17~ & ~0.63~(X)~ & ~0.77~(M)~  \\
~2.89~ & ~1.72~ & ~1.00~ & ~0.17~ & ~0.74~(X)~ & ~0.76~(M)~  \\
~2.83~ & ~1.66~ & ~1.00~ & ~0.17~ & ~0.89~(X)~ & ~0.76~(M)~  \\
\hline
\end{tabular} 
\label{tab:occ_dop}
\end{table}

Table~\ref{tab:occ_dop} shows the occupation of the bands, the LEV and the corresponding eigenvectors of the charge and spin fluctuations.
We find that upon the hole doping the total density $n_{\rm total}$ is reduced by diminishing the occupation of the $d_{z^2}$ band, while the occupation of the other two bands remains unchanged.
In particular, the $d_{x^2-y^2}$ stays half-filled and Mott-insulating for all considered levels of the doping, which consequently leads to a nearly unchanged strength (spin LEV) of the leading \mbox{C-type} (${{\bf q}=\text{M}}$) AFM fluctuations that stem from this band. 
This fact explains the intrinsic magnetic ground state that was detected experimentally for various superconducting infinite-layer nickelates irrespective of the rare earth ion or doping~\cite{fowlie2022intrinsic}.

\begin{figure}[t!]
\includegraphics[width=1\linewidth]{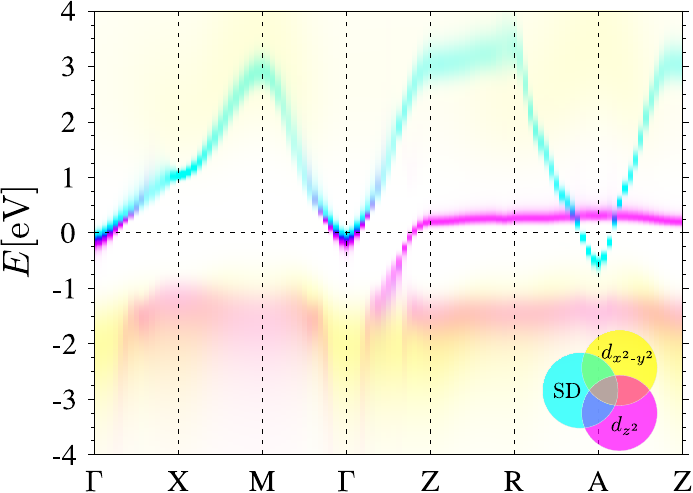}
\caption{The momentum-resolved electronic spectral function. The result is obtained for the $d_{z^2}$, $d_{x^2-y^2}$, and SD bands (see the inset for the color code) for ${\mu_{xy}}$ and  ${n_{\rm total}=2.83}$ near the CDW instability (charge ${\text{LEV}=0.89}$) along the high-symmetry path in the BZ.
\label{fig:A_dop}}
\end{figure}

We find that the mechanism of the formation of the CDW ordering also remains the same as in the half-filled case.
The only difference is that the reduction of $n_{z^2}$ is achieved here by the hole doping instead of tuning $\mu^{\rm DC}$.
As can be found from Table~\ref{tab:occ_dop}, reducing the occupation of the $d_{z^2}$ band consequently enhances the strength of the charge fluctuations, since they originate from the intraband electronic scattering within the $d_{z^2}$ band.
We observe that the momentum-dependent renormalization of the spectral function 
upon doping is qualitatively the same as upon changing $\mu^{\rm DC}$.
In Fig.~\ref{fig:A_dop} the momentum-resolved spectral function is shown in the vicinity of the CDW instability (charge ${\text{LEV}=0.89}$) induced by the doping (${n_{\rm total}=2.85}$).
We again notice the splitting of the flat part of the $d_{z^2}$ band at ${k_{z}=\pi}$ into the two parts and the hybridization of its lower part with the lower Hubbard $d_{x^2-y^2}$ band. 
We also see that near the CDW instability the upper part of the flat band appears above the Fermi energy and thus does not participate in the two-particle scattering that results in the CDW instability.
In turn, the $\Gamma$ point of the $d_{z^2}$ band is shifted to lower energies below $E_{F}$ and forms the electron pocket.

Interestingly, the renormalization of the spectral function upon hole doping found in \mbox{D-TRILEX} is different from the one obtained within the DFT+sicDMFT scheme~\cite{PhysRevX.10.041002}.
The latter predicts a rather uniform shift of the bands toward higher energies.
In particular, DFT+sicDMFT the flat part of the $d_{z^2}$ band does not split and moves from below to above the Fermi energy.
The $\Gamma$ point also shifts to positive energies, which removes the electron pocket from the Fermi level.
Pinning the electron pocket to the Fermi energy upon doping observed in \mbox{D-TRILEX} is, therefore, an important effect of nonlocal electronic correlations, since this pocket plays a crucial role in the formation of the CDW ordering.
In particular, this pinning explains why the CDW ordering vector ${{\bf q}_{\rm CDW}=\text{X}}$ remains unchanged for different doping levels (Table~\ref{tab:occ_dop}).
We note that the spectral function in Fig.~\ref{fig:A_dop} is plotted at a smaller charge LEV than the one in the right panel of Fig.~\ref{fig:A} due to convergence issues while approaching the CDW instability in the doped case. 
For this reason, the flat part of the $d_{z^2}$ band is not yet aligned with the vHS of the SD band, and the electron pocket at the $\Gamma$ point is smaller than the one in the right panel of Fig.~\ref{fig:A}.

\begin{figure}[t!]
\includegraphics[width=1\linewidth]{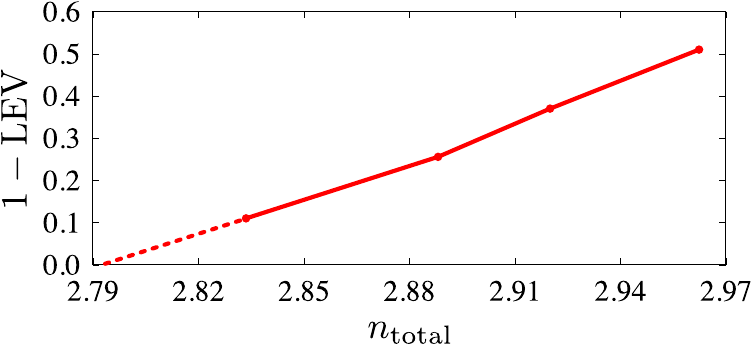}
\caption{The evolution of the charge LEV as a function of the hole doping. To estimate the critical value of the doping we plot the \mbox{(1-LEV)} value that becomes zero at the transition point. The results are obtained in the presence of the coupling. According to a linear fit performed for the two lowest points the critical density for the CDW phase transition in the presence of phonons is ${n_{\rm total}\simeq2.79}$.
\label{fig:le}}
\end{figure}

The critical density for the CDW transition can be obtained by looking at the evolution of the charge LEV as a function of doping. 
In Fig.~\ref{fig:le} we plot the difference of the LEV from unity (${1-\text{LEV}}$) as a function of the total electronic density $n_{\rm total}$.
By extrapolating the ${1-\text{LEV}}$ value to zero we get the ${n_{\rm total}\simeq2.79}$ value for the critical density for the CDW phase transition.
It is important to point out that the CDW instability that originates due to the Coulomb interaction (not via the electron-phonon mechanism) typically does not exhibit a strong dependence on temperature~\cite{PhysRevB.95.115149, stepanov2021coexisting, vandelli2024doping}.
This allows us to speculate that the dip in the superconducting dome observed in the hole doped NdNiO$_2$ at the density ${n_{\rm total}\simeq2.80}$~\cite{PhysRevLett.125.027001, PhysRevLett.125.147003} may arise due to a competition of the superconductivity with strong CDW fluctuations.
A similar conclusion was made in the recent experimental work~\cite{ji2022rotational}, where the rotational symmetry breaking observed in superconducting Nd$_{0.8}$Sr$_{0.2}$NiO$_{2}$ films was associated with the charge order.
We also wish to point out, that the doping level at which these measurement were performed is similar to the critical density for the CDW phase transition obtained in the current work.

\begin{figure*}[t!]
\includegraphics[width=1\linewidth]{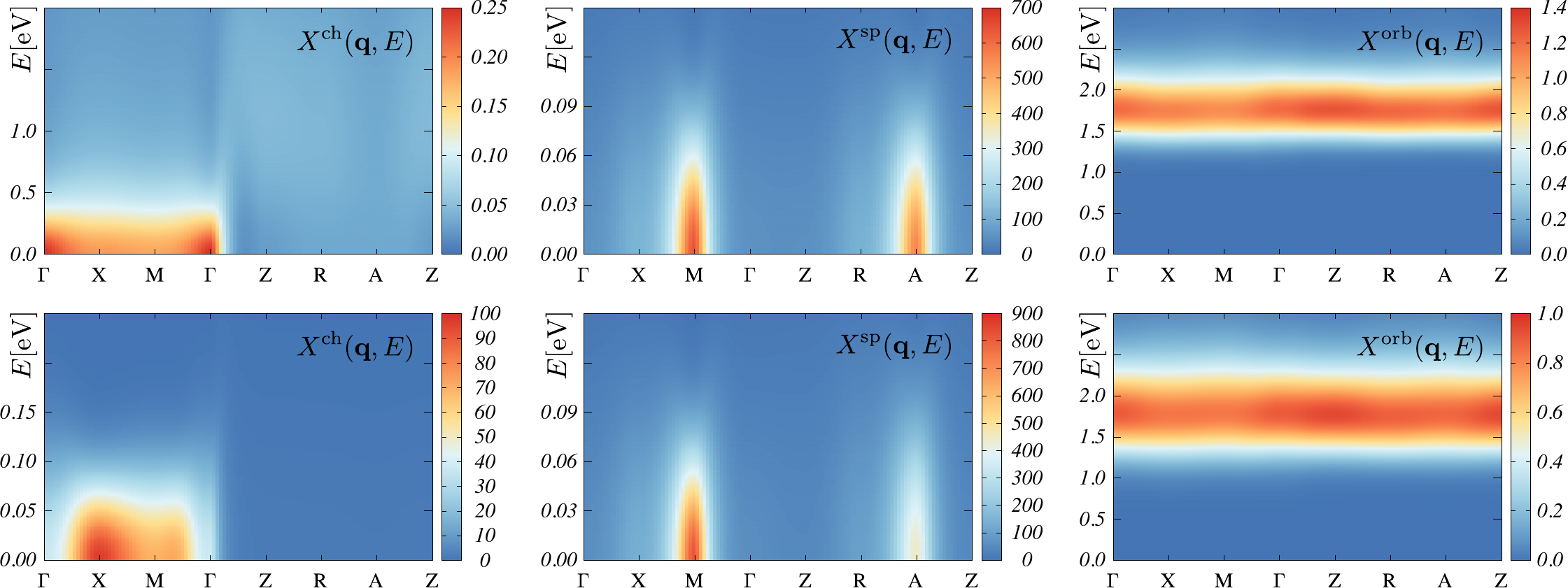}
\caption{Momentum-resolved spectral functions for collective electronic fluctuations. Charge (left panels), spin (middle panels) and orbital (right panels) spectral functions (susceptibilities) calculated as a function of real energy $E$. Results are obtained for ${\mu_{xy}}$ for the correlated $d_{z^2}$ and $d_{x^2-y^2}$ bands along the high-symmetry path in the BZ. Calculation are performed at half filling (top row) and at ${n_{\rm total}=2.83}$ (bottom row).
\label{fig:Xlatt_q}}
\end{figure*}

In Fig.~\ref{fig:Xlatt_q} we plot the spectral functions (susceptibilities) for collective charge (left column), spin (middle column) and orbital (right column) fluctuations as a function of real energy $E$ along the high-symmetry path in the BZ.
The result is obtained for the most consistent DC correction $\mu_{xy}$ at half-filling (top row) and in the vicinity of the CDW instability for ${n_{\rm total}\simeq2.83}$ (bottom row).
These results visualise the evolution of the collective electronic fluctuations with doping discussed above. 
At stoichiometry the charge fluctuations are negligibly small (top left panel) and dominated by the low energy (static) contribution. 
The spin fluctuations are also static and display the leading ${{\bf q}=\text{M}}$ (C-type) and the subleading ${{\bf q}=\text{A}}$ (G-type) AFM modes that have a comparable intensity (top middle panel).
We note that dispersive spin excitations (magnons) are not visible in this plot because the calculations are performed close to the magnetic instability, and thus almost all spectral weight is moved to the ordering AFM modes. 
Instead, the orbital fluctuations are dynamic with the characteristic energy of ${E\simeq1.5-2.0}$\,eV but are rather small and nearly momentum-independent (top right panel).

Upon approaching the CDW instability the charge spectral function reduces to a two-peak structure with the main mode at ${{\bf q}=\text{X}}$, which will lead to the CDW ordering upon the transition, and a small satellite peak at ${{\bf q} = (\pi/3, \pi/3, 0)}$ (bottom left panel).
In turn, the spin spectral function still reveals two AFM modes (bottom middle panel).
However, the subleading mode in the doped case is more suppressed with respect to the leading compared to the half-filled case.
The competition between the C- and G-type AFM ordering that at large doping levels reduces to the C-type one has also been found in the previous DMFT calculation~\cite{LEONOV2021160888}. 
Finally, orbitals fluctuations remain qualitatively unchanged upon doping, but become slightly weaker and broader in energy.  

\section{Discussion}

In this work we investigated the effect of collective electronic fluctuations in the infinite-layer NdNiO$_2$ at stoichiometry and upon hole dpoing.
We have shown that the electronic correlations that lead to the CDW and magnetic instabilities in this system are orbital-dependent.
We have found that the strong CDW fluctuations are related to electronic correlations within the $d_{z^2}$ orbital.
Upon doping, these fluctuations drive the material toward the CDW ordered phase when considering only possible particle-hole instabilities. The mechanism of this transition relies on the strong momentum-dependent renormalization of the electronic spectral function.
This renormalization consists of splitting and moving one part of the flat region of the $d_{z^2}$ band from below to above the Fermi energy, where it aligns in energy with the vHS of the SD band.
The other part of the flat band is moved in the opposite direction, where it hybridizes with the lower Hubbard $d_{x^2-y^2}$ band.
In turn, the $\Gamma$ point of the hybridized $d_{z^2}$ and SD bands is shifted toward smaller energies, which enlarges the electron pocket around the vicinity of the $\Gamma$ point.
As a result, the flat part of the band does not take part in the formation of the CDW phase, that instead occurs due to the nesting of the Fermi surface in the $(k_x,k_y,\pi/8)$ plane.

We argue that this complex momentum-dependent renormalization of the electronic spectral function, which is associated with a rather large redistribution of the electronic density between the orbitals, is unlikely to happen in the stoichiometric case.
This means that the formation of the CDW ordering cannot be found in the pristine NdNiO$_2$, which is also confirmed by the most recent experiments~\cite{https://doi.org/10.1002/smll.202304872, parzyck2023, hayashida2024investigation}.
Instead, we demonstrate that the same renormalization can be obtained upon hole doping leading to the CDW phase transition at a critical density ${n_{\rm total}\simeq2.79}$.
Remarkably, we find that the nonlocal electronic correlations pin the electron pocket of the Ni-$d_{z^2}$ band, which is responsible for the formation of the CDW ordering, to the Fermi energy even upon hole doping.
This observation challenges the single-band picture of the electronic correlations in nickelates, which partially originates from the DMFT prediction that the electron pocket becomes unimportant upon doping as it shifts well above the Fermi level (see, e.g., Ref.~\cite{PhysRevX.10.041002}).

We also find that the strong magnetic fluctuations originate from the intraband electronic correlations within a completely different, namely the $d_{x^2-y^2}$, orbital.
Our calculations show that the electronic correlations redistribute the density between the orbitals in such a way that the $d_{x^2-y^2}$ band remains half-filled.
Remarkably, the filling of the $d_{x^2-y^2}$ band remains unchanged upon hole doping, which, consequently, results in a nearly unchanged strength of magnetic fluctuations that consist of the leading C- and the subleading G-type AFM modes.
These results show that both, CDW and magnetic fluctuations are strong in the hole doped regime of NdNiO$_2$ and can thus affect superconductivity in this compound.

\section{Methods}

The three-orbital Hubbard-Holstein-Kanamori Hamiltonian~\eqref{eq:H_latt} that models the NdNiO$_2$ compound is solved using the many-body diagrammatic \mbox{D-TRILEX} approach~\cite{PhysRevB.100.205115, PhysRevB.103.245123, 10.21468/SciPostPhys.13.2.036}.
This method is a dual version~\cite{PhysRevB.77.033101, PhysRevB.79.045133, PhysRevLett.102.206401, Rubtsov20121320, PhysRevB.90.235135, PhysRevB.93.045107, PhysRevB.94.205110, PhysRevB.100.165128, PhysRevB.102.195109} of the TRILEX approach~\cite{PhysRevB.92.115109, PhysRevB.93.235124, PhysRevB.96.104504, PhysRevLett.119.166401}, where the nonlocal electronic correlations are taken into account by means of the diagrammatic expansion constructed on the basis of an interacting reference problem.
The latter is chosen in such a way that it can be solved numerically exactly, e.g., by using the continuous time quantum Monte Carlo (CT-QMC) solvers~\cite{PhysRevB.72.035122, PhysRevLett.97.076405, PhysRevLett.104.146401, RevModPhys.83.349}.
This sets certain limitations on the form of the reference system.
In particular, in the multi-band case the advanced CT-QMC solvers are usually restricted to a static interaction.
For this reason, in this work we perform \mbox{D-TRILEX} calculations based on the DMFT reference impurity problem that involves only the static interaction $U_{l_1 l_2 l_3 l_4}$.
The frequency-dependent interaction ${U^{\rm ph}_{ll'll'}(\omega)}$ that originates from the electron-phonon coupling is treated diagrammatically in the same way as one would account for the nonlocal Coulomb interaction~\cite{PhysRevB.100.205115, PhysRevB.103.245123, 10.21468/SciPostPhys.13.2.036}.
The impurity problem is solved using the \textsc{w2dynamics} package~\cite{WALLERBERGER2019388}.
All calculations are performed for ${N_{k}=32^3}$ number of ${\bf k}$-points in the BZ.
The local and momentum-resolved electronic, charge, spin and orbital spectral functions are obtained from the corresponding Matsubara Green's functions and susceptibilities via analytical continuation using the maximum entropy method implemented in the \textsc{ana\_cont} package~\cite{kaufmann2021anacont}. 
The charge and spin susceptibilities are obtained using \mbox{D-TRILEX}, as described in Ref.~\onlinecite{10.21468/SciPostPhys.13.2.036}.\\

{\bf Data Availability.} The data that support the findings of this work are available from the corresponding author upon reasonable request. 

{\bf Code Availability.} Many-body calculations have been performed using the implementation of the \mbox{D-TRILEX} method~\cite{10.21468/SciPostPhys.13.2.036}. 
The \mbox{D-TRILEX} code is available from the corresponding author upon reasonable request.

{\bf Funding.} A.I.L. acknowledges the support by the DFG through FOR 5249-449872909 (Project P8) and by the European Research Council via Synergy Grant 854843-FASTCORR.
E.A.S. acknowledges the help of the CPHT computer support team.

{\bf Author contributions.} All authors discussed the results and contributed to the preparation of the manuscript.
 
{\bf Competing Interests.} The Authors declare no Competing Financial or Non-Financial Interests.

\bibliography{Ref}

\end{document}